\documentclass[a4,12pt]{article}
\usepackage[dvipdfmx]{graphicx,hyperref}
\usepackage{graphicx}
\usepackage{epsf}
\usepackage{psfig}
\usepackage{natbib}
\usepackage{authblk}

\begin{document}
\title{Comments on "Imaging Reanalyses of EHT Data"}
\author{Makoto Miyoshi}
\affil{National Astronomical Observatory, Japan, 2-21-1, Osawa, Mitaka, Tokyo, Japan, 181-8588, e-mail: makoto.miyoshi@nao.ac.jp}
\author{Yoshiaki Kato}
\affil{Japan Meteorological Agency, 3-6-9 Toranomon, Minato City, Tokyo 105-8431, Japan, e-mail: yoshi\_kato@met.kishou.go.jp}
\author{Junichiro Makino}
\affil{Department of Planetology, Kobe University, 1-1 Rokkodaicho, Nada-ku, Kobe, Hyogo 650-0013, Japan, e-mail: makino@mail.jmlab.jp}

\maketitle

On June 14 2022, the EHT collaboration (hereafter EHTC)  made the
web page\footnote{
\url{https://eventhorizontelescope.org/blog/imaging-reanalyses-eht-data}
}
with the title "Imaging Reanalyses of EHT Data", in which they made
comments on our recent paper\cite{Miyoshietal2022}
published in the Astrophysical Journal.

The EHTC's comment on our paper is the following:

\begin{quote}
  The EHT images of M87 are among the most vetted interferometric
  images ever published \citep{EHTC2019IV,EHTC2019VI}. Four
  independent analyses
  \citep{Arrasetal2022, CarilliandThyagarajan2022,LockhartandGralla2022, Pateletal2022arXiv} have reconstructed the ring-like structure of M87, employing a diverse set of techniques.  These efforts complement the three imaging and two modeling techniques in the 2019 EHTC papers presenting the first M87 results.  Furthermore, the EHTC and its members have published two additional papers, employing newly developed and independent techniques, that confirm the original results \citep{EHTC2021VII, SunandBouman2021}.  As an ensemble, these multiple independent techniques and efforts are robust against algorithmic bias, parameter selection, or human bias.

  Our team has determined that a new re-analysis\citep{Miyoshietal2022} is based on a flawed understanding of EHTC data and its methods, leading to
  erroneous conclusions.  Ring-like structures are unambiguously recovered under a broad range of imaging assumptions, including field of view.  Additionally, large-scale jet structures are unconstrained by this high-resolution data.
\end{quote}
(N.B. The reference style in the above have been changed in order to match our list of reference.)

In the above, EHTC's claim consists of the following five points:

\begin{description}
\item{(a)} The EHT images of M87 are among the most vetted interferometric
images ever published \citep{EHTC2019IV,EHTC2019VI}.

\item{(b)} Four independent analyses \citep{Arrasetal2022,
    CarilliandThyagarajan2022,LockhartandGralla2022, Pateletal2022arXiv}
have reconstructed the ring-like structure of M87, employing a
diverse set of techniques.

\item{(c)} The EHTC and its members have published two additional papers,
employing newly developed and independent techniques, that confirm the
original results \citep{EHTC2021VII, SunandBouman2021}.

\item{(d)} The EHTC has determined that a new re-analysis \citep{Miyoshietal2022}  is based on a flawed understanding of EHTC data and its methods.

\item{(e)} Ring-like structures are unambiguously
recovered under a broad range of imaging assumptions, including field
of view. Additionally, large-scale jet structures are unconstrained
by this high-resolution data.
\end{description}

Let us investigate all points summarised above in detail.\\

Point (a) is the subjective claim of the EHTC without any supporting data or fact. In our paper we have demonstrated that this claim is not true.\\

Concerning point (b), it  is important whether    the analyses in four papers are really  independent and supporting the EHTC result  or  not. So we have carefully studied the "Four independent analyses" and have found that, one of them, \citet{CarilliandThyagarajan2022} actually obtains the result very similar to ours, using the imaging algorithm similar to what we have used (the so-called hybrid mapping).
In Figure 4 of \cite{CarilliandThyagarajan2022} it is shown that an annulus appears  when they use a ring or a disk as initial models, but it is also shown that something else appears when they use a single point, double points or extended Gaussian as initial models. Actually the image they have reported for the ring initial model is close to what we have obtained using the EHTC ring as the initial model (figure 20 of \cite{Miyoshietal2022}), and more importantly, the image they have reported for the single point initial model is very close to our result (figure 7 of \cite{Miyoshietal2022}). Therefore, one of the four papers the EHTC listed as "have reconstructed the ring-like structure" prove the validity of our result as well.  Therefore, EHTC's statement "new re-analysis(Miyoshi et al., 2022) is based on a flawed understanding of EHTC data and its methods, leading to erroneous conclusions." doesn't make sense.

Concerning the other three papers, 
\citet{LockhartandGralla2022} started from a ring, 
\citet{Arrasetal2022} used essentially the same method as that of EHTC,
and
\citet{Pateletal2022arXiv} used the EHTC software itself.  
Therefore
these three papers cannot be regarded as "independent analyses".\\

Point (c) does not really add anything new to point (a), since the methods
they used in these papers have the same problems as those in their original papers.
\citet{EHTC2021VII} tried to determine the polarization but the imaging software is the same as that used in Paper IV(SMILI). \citet{SunandBouman2021} used a machine-learning technique,
and apparently their training input images are all compact. Thus, most
likely the neural network of \cite{SunandBouman2021} is trained to find compact structure, even when the data actually contain emissions from a wider region.\\

In point (d), the EHTC claimed our re-analysis is based on "a flawed understanding of EHTC data and its methods." However, the EHTC did not make clear where our understanding is flawed. Moreover, as we have stated above,
one of the "Four independent analyses", \citet{CarilliandThyagarajan2022}, have actually obtained the result very similar to ours. Since the method and results of \cite{CarilliandThyagarajan2022} are quite close to ours, 
EHTC should make clear what is "a flawed understanding of their data
and methods", not only for our work but also for \citet{CarilliandThyagarajan2022}.\\

In point (e), the EHTC claimed that Ring-like structures are recovered
under a broad range of imaging assumptions, including the field of view.
However, the actual fields of view set by the EHTC are limited to a very narrow area.
In \cite{Miyoshietal2022}, we have calculated and shown the range of the field of view (FOV) over which the data contains information.
The FOV settings of the EHTC are two orders of magnitude narrower than ours.
For example, in the EHTC Paper IV (\cite{EHTC2019IV}), the EHTC stated:

\begin{quote}
We (= EHTC) limited the cleaning windows to image only the
compact structure ($<100\rm \mu as$) in order to prevent CLEAN from
adding larger-scale emission features that are poorly constrained by
the lack of short EHT baselines. (section 6.2.1 4th paragraph)
\end{quote}

and

\begin{quote}
Images that meet the threshold for the Top Set are outlined in green (see
Section 6.3.1). (Figure 6 caption).
\end{quote}
In figure 6 of \cite{EHTC2019IV}, only the fields of view with diameter 60, 70 and 80 $\mu
\rm as$
are marked as "Top Sets", representing the combination of parameters which  produced acceptable images on the synthetic data. Thus, the EHTC's statement in point (e), ring-like structures are recovered for a broad range of field of view, is hard to accept. Of course, the EHTC can argue that the other images are not acceptable but still "ring-like".  However, it is important to note that, from figure 6, the images with the mask of 100 $\mu \rm as$ is noticeably different from those obtained using masks with smaller diameters. Actually, the EHTC discussed on this issue in section 6.3.2:
\begin{quote}
  A common feature of all four geometric models in the training set is
  that they restrict the compact emission to a region spanning $\sim
  70\rm~\mu as$. One possible concern is that the resultant Top Set parameter combinations may then fail to reconstruct images of sources that
  have compact components extending over larger regions. However, such
  bright compact components would introduce variations over time in
  the visibilities that could not be fit with an overly restricted
  FOV.
\end{quote}
In the above the EHTC argued that if compact components exist outside the 70 $\mu \rm as$ diameter mask, it should generate large residuals since the image limited to small FOV cannot express that components. Unfortunately, the EHTC have not verified this claim with simulated data in published papers. They admitted that when a large mask is used for compact simulated data, resulted images would contain "larger-scale emission features that are poorly constrained by the lack of short EHT baselines" (see 4th paragraph in section 6.2.1). In other words, the CLEAN results which contain larger-scale emission features can fit the observed visibility without such larger-scale emission features, since, as the EHTC stated, they "are poorly constrained by the lack of short EHT baselines".  Mathematically, this means that the CLEAN result which does not contain larger-scale emission features can fit the observed visibility with such larger-scale emission features. We would strongly suggest the EHTC should try various models with "compact components extending over larger regions" and check if they can be actually excluded or not.\\

In conclusion, all of the five points raised by the EHTC are subjective and 
unsubstantiated claims. Thus they do not prove the correctness of the result of EHTC. Sincerely we hope that the EHTC will publish, not a collection of unsubstantiated claims, but a discussion based on scientific arguments.

Otherwise, they should retract the statement of "new re-analysis(Miyoshi et al., 2022) is based on a flawed understanding of EHTC data and its methods, leading to erroneous conclusions."

\def\pasj{PASJ}
\def\apj{ApJ}
\def\aj{AJ}
\def\apjl{ApJL}
\def\apjs{ApJS}
\def\MN{MN}
\def\mnras{MN}
\def\mn{MN}
\def\jcp{J. Comp. Phys.}


\newcommand{\noopsort}[1]{} \newcommand{\printfirst}[2]{#1}
  \newcommand{\singleletter}[1]{#1} \newcommand{\switchargs}[2]{#2#1}

\end{document}